\documentclass{aa}
\usepackage{graphicx}
\usepackage{txfonts}
\begin{document}
   \title{The effects of ram-pressure stripping on the internal kinematics of simulated spiral galaxies}

   \author{T. Kronberger \inst{1,2}
          \and
           W. Kapferer \inst{1}
          \and
           S. Unterguggenberger \inst{1}
          \and
           S. Schindler \inst{1}
          \and
           B. L. Ziegler \inst{3}
           }

   \offprints{T. Kronberger, \email{Thomas.Kronberger@uibk.ac.at}}

   \institute{Institute for Astro- and Particle Physics, University
   of Innsbruck, Technikerstr. 25, A-6020 Innsbruck, Austria
   \and
   Max-Planck-Institut f\"ur Extraterrestrische Physik, Giessenbachstrasse, D-85748 Garching bei M\"unchen, Germany
   \and
   European Southern Observatory, Karl-Schwarzschild-Strasse 2, D-85748 Garching bei M\"unchen, Germany
   }

   \date{-/-}


  \abstract
   {}
   {We investigate the influence of ram-pressure stripping on the internal gas kinematics
   of simulated spiral galaxies. Additional emphasis is put on the question of how the resulting
   distortions of the gaseous disc are visible in the rotation curve and/or the full 2D velocity
   field of galaxies at different redshifts.}
   {A Milky-Way type disc galaxy is modelled in combined N-body/hydrodynamic simulations with prescriptions for
   cooling, star formation, stellar feedback, and galactic winds. This model galaxy moves through
   a constant density and temperature gas, which has parameters similar to the intra-cluster medium
   (ICM). We study five different configurations, in which the direction of the ram pressure on the gaseous disc
   varies from face-on to edge-on. Rotation curves (RCs) and 2D velocity fields of the gas are extracted
   from these simulations in a way that follows the procedure applied to observations of distant, small, and faint
   galaxies as closely as possible.}
   {We find that the appearance of distortions of the gaseous disc due to ram-pressure stripping depends on the
   direction of the acting ram pressure. In the case of face-on ram pressure, the distortions mainly appear in the outer parts
   of the galaxy in a very symmetric way. This distinction between undisturbed inner part and symmetrically disturbed outer
   regions is visible in both the rotation curve and the 2D velocity field. In contrast, in the case of edge-on ram pressure
   we find stronger distortions. In particular, a mismatch between kinematic centre and the centre of the stellar disc becomes
   observable. In the RC, there are basically two features visible. The RC is asymmetric as the kinematic centre does
   not coincide with the optical centre of the galaxy. Secondly, the outer parts of the RC are declining. The 2D velocity field
   also shows signatures of the interaction in the inner part of the disc. At angles smaller than 45$^{\circ}$ between
   the ICM wind direction and the disc, the velocity field asymmetry increases significantly compared to larger angles.
   Compared to distortions caused by tidal interactions, the effects of ram-pressure stripping on the velocity field are
   relatively low in all cases and difficult to observe at intermediate redshift in seeing-limited observations.}
   {}


   \maketitle
%

\section{Introduction}

It has been pointed out ever since the late 1970's that there is an
evolution of galaxy cluster members with redshift. The fraction of
star-forming and post-star-forming systems, for example, has been
proven to increase significantly with $z$ (Dressler et al. 1987,
1999). Consistently, Butcher \& Oemler (1978) reported a strong
evolution from bluer to redder colours in cluster galaxies,
detecting an excess of blue objects at $z$=0.5 with respect to lower
redshift systems (this observational finding is usually referred to
as 'Butcher -- Oemler effect'). Various physical mechanisms were
proposed which act on the star formation activity and on the
morphology of galaxies in clusters besides the hierarchical assembly
of structures (e.g. ram-pressure stripping: Gunn \& Gott 1972 --
``harassment'': e.g. Moore et al. 1998 -- ``strangulation'' or
``starvations'': e.g. Larson et al. 1980). We still need to
understand the specific importance of each of the proposed
processes. For that purpose it is crucial to identify and
disentangle different interactions, also at intermediate and high
redshift. One possibility is to study the stellar population of the
galaxies which, however, depends heavily on the current star
formation activity. The time-dependent star formation encompasses
several non-linear processes, which are partly poorly understood.
Another method is to study the total gravitational potential of
galaxies and its possible distortions due to interactions via their
internal kinematics. Technically it became feasible to observe the
full 2D velocity field (VF) of local galaxies in optical wavebands
using integral field units (IFUs) such as SAURON (e.g. Ganda et al.
2006) or Fabry-Perot interferometry (e.g. Chemin et al. 2006;
Garrido et al. 2002). For intermediate and high redshift galaxies,
however, there are few observational studies of 2D velocity fields
available and many studies still rely on rotation curves (RCs) from
slit spectroscopy (e.g. Weiner et al. 2006; Moran et al. 2007;
B\"ohm \& Ziegler 2007). To identify distortions and irregularities
in the velocity fields is in both cases critical. A desirable goal
would be to disentangle different interaction processes due to their
influence on the velocity field. Numerical simulations offer a
possibility to study individually the effects of the various
processes. Kronberger et al. (2006, 2007) investigated in this way
the effects of galaxy-galaxy mergers and of tidal interactions
between galaxies on the internal kinematics of model galaxies. Those
simulations were also used to identify possible observational biases
in observations of the velocity field of distant galaxies (Kapferer
et al. 2006; Kronberger et al. 2007). Although dependencies on the
viewing angle and on the spatial resolution have been found, it was
shown that tidal interactions mainly introduce nonaxisymmetric and
nonbisymmetric features (see also Rubin et al. 1999).

Ram-pressure stripping is another important cluster-specific
interaction process. The space between the galaxies in a galaxy
cluster is filled with a hot ($\sim10^8$ K), thin ($\sim10^3$
ions/m$^3$) plasma, the so called intra-cluster medium (ICM). This
plasma exerts a pressure on the inter-stellar medium (ISM), which
can remove gas from the disc if the force due to ram-pressure
stripping exceeds the restoring gravitational force of the galaxy.
This mass loss due to ram-pressure stripping was modelled by many
groups (e.g. Abadi et al. 1999; Vollmer et al. 2001; Roediger \&
Hensler 2005; Quilis et al. 2000; Mori \& Burkert 2000; Toniazzo \&
Schindler 2001; Schulz \& Struck 2001; and J{\'a}chym et al. 2007).
Recently also the effects of the external ram pressure on the star
formation rate of the stripped galaxy were studied (Kapferer et al.
2008, Kronberger et al. 2008). Here we study the influence of
ram-pressure stripping on the internal gas kinematics of simulated
spiral galaxies.

This interaction mechanism has little influence on the stellar
structure and could, therefore, be hidden in a galaxy that has a
regular appearance when only direct imaging is used to search for
interaction processes. However, it has a strong effect on the gas
clouds within a galaxy disk disrupting its rotational motion, which
can be traced by emission lines originating in these gas clouds. If
spectra of observed galaxies have sufficient signal in the
continuum, absorption lines can be used to also derive
\textit{stellar} rotation curves. Discrepancies between these and
the gas motion are then strong indications of ram pressure. The
velocity field of faint distant galaxies is in most cases measured
only by emission lines (e.g. Ziegler et al. 2002, Bamford et al.
2005, Metevier et al. 2006) so that one needs a very thorough
understanding of the effect of ram-pressure stripping.

Observable signatures of ram-pressure stripping were also studied
numerically by Roediger \& Br\"uggen (2006, 2007) using Eulerian
hydrodynamics. They focussed on the gas distribution in the disc and
in the wake and thereby tried to deduce information on the
interaction geometry. The velocity field in the wake has also been
investigated. They have, however, not considered the velocity field
in the disc in detail, as we do it in the present work. In a series
of papers Vollmer et al. have used sticky particle simulations to
model specific observed systems (Vollmer et al. 2000, 2003, 2006,
2008). They have tried to match observed HI velocity fields and gas
distributions to infer information on the interaction scenario.
Consequences of ram-pressure stripping for observable ICM properties
are discussed in Schindler \& Diaferio (2008) and references
therein. This present work goes beyond these previous studies by
systematically investigating the effects of ram-pressure stripping
observable with rotation curves from optical long-slit spectroscopy
and 2D velocity fields at different redshifts.

We also study whether the deviations from a normal rotation caused
by ram-pressure stripping have a different form than in the case of
other interactions, so that they could be taken as an indicator of
what mechanism may be acting on a galaxy. In the cluster and group
environment it is expected that ram-pressure stripping is one of the
main drivers for evolutionary effects on galaxies. It could be
responsible for the global suppression of star formation in cluster
galaxies (either through an initial star burst triggered by gas
compression, as found e.g. by Kronberger et al. (2008) and Kapferer
et al. (2008), or through the subsequent removal of gas from the
disk). On the other hand, it cannot be responsible for structural
changes needed for a complete transition from one galaxy type to
another.

The paper is organised as follows: in Sect. 2 we present the
simulation setup used for this work. In Sect. 3 the extraction of
the observable quantities (i.e. the rotation curves and the 2D
velocity fields) are described. The results are presented in Sect. 4
and subsequently discussed with respect to previous observational
and numerical work in Sect. 5. We end with a summary of the main
conclusions in Sect. 6.

\section{Simulations}

The simulations were carried out with the N-body/SPH code GADGET-2
developed by V. Springel (see Springel 2005 for details). This code
treats the gas of the galaxies and the ICM hydrodynamically via
smoothed particle hydrodynamics (SPH, Gingold \& Monaghan 1977; Lucy
1977) while the collisionless dynamics of the dark matter and the
stellar component is calculated using an N-body technique.
Prescriptions for cooling, star formation (SF), stellar feedback,
and galactic winds are included as described in Springel \&
Hernquist (2003). The mass loss of the galaxy due to galactic winds,
$\dot{M}_w$, is assumed to be proportional to the star formation
rate $\Psi_{SFR}$, i.e. $\dot{M}_w=\eta\Psi_{SFR}$ with $\eta=2$,
consistent with the observations of Martin (1999). Additionally, the
wind contains a fixed fraction $\chi$ of the supernova energy, which
is assumed to be $\chi$=0.25, as in Springel \& Hernquist (2003).

The initial conditions were built according to Springel et al.
(2005), based on the analytical work of Mo et al. (1998). The model
galaxy, which we use, represents a Milky-Way type spiral galaxy. The
total mass of the model galaxy is 1.09$\times$10$^{12}$ $M_{\sun}$,
where the initial total gas mass is 6.8$\times$10$^{9}$ $M_{\sun}$.
The initial disc scale length is $r_d \sim 3.3$ kpc. Compared to
previous work (Kapferer et al. 2005, 2006; Kronberger et al. 2006,
2007), where we used a similar sized galaxy, we have significantly
increased the number of particles, increasing the mass resolution of
the simulation. Important numerical quantities, such as the particle
numbers, are summarised in Table \ref{galaxyproperties_resolution}.
For the intra-cluster medium we use one million SPH particles, which
we distribute homogeneously over a volume of 1 Mpc$^3$ with a mass
density of $1\times10^{-28}$ g/cm$^3$ and a constant temperature of
3 keV ($\sim 3.6\times10^{7}$ K). Such a simplified ICM distribution
allows the effects of ram-pressure stripping to be studied in a
clean way, i.e. with as little degeneracies of different effects as
possible. Additional effects from varying density and temperature in
the ambient medium will be investigated in an upcoming work.

\begin{table}
\begin{center}
\caption[]{Particle numbers and mass resolution used for the model
galaxy.}
\begin{tabular}{c c c}
\hline \hline & particle number & mass resolution \cr & &
[$M_{\sun}$/particle] \cr\hline Dark matter halo & 300000 &
$3.5\times10^{6}$ \cr Disk collisionless & 200000 &
$1.0\times10^{5}$ \cr Gas in disk & 200000 & $3.4\times10^{4}$ \cr
ICM & 1000000 & $1.4\times10^{6}$\cr\hline
\end{tabular}
\label{galaxyproperties_resolution}
\end{center}
\end{table}

Starting from these initial conditions we calculate the evolution of
the model galaxy moving with a constant velocity of 1000 km/s
through the ambient medium for 1 Gyr. We calculated five different
interaction geometries, where we varied the angle between the ICM
wind direction and the plane of the disc from 0$^{\circ}$ to
90$^{\circ}$. The edge-on and face-on stripping configurations are
the two extreme cases where the perturbation acts either
perpendicular to the galactic disc or purely in the plane of the
gaseous disc. We study these two cases in detail. The other three
simulations, where this angle is 23$^{\circ}$, 45$^{\circ}$, and
68$^{\circ}$, respectively, are used to study how the asymmetry of
the velocity field depends on the ICM wind direction.

\section{Extraction of realistic rotation curves and 2D velocity fields}

In order to be able to compare our results to observations of
distant, small, and faint galaxies we extract rotation curves and 2D
velocity fields as described in Kronberger et al. (2006, 2007) in a
way which closely resembles the observational approach (e.g. Ziegler
et al. 2003). For the construction of realistic 2D velocity fields,
we project all gas particles onto a Cartesian, equally spaced grid.
The spacing is chosen such that the spatial resolution corresponds
to the angular resolution of current state-of-the-art observations.
We adopt an angular resolution typical for IFU or FPI observations,
namely 0.4" (e.g. Chemin et al. 2006). The angular resolution of
SAURON, for example, would be 0.3" or 0.9". We calculate the
physical resolution according to the given angular resolution for a
particular redshift using the concordance cosmological model. The
velocity field of the galaxy is binned using this spatial
resolution.

For observations of galaxies at intermediate and high redshift
seeing plays a crucial role, as it typically exceeds the angular
resolution of the instrument. To simulate seeing effects on the
velocity-field measurements we apply a convolution with a Gaussian
point spread function. A value of 0.8" for the FWHM of the Gaussian
seeing was adopted, which is a typical value for ground based
observations (see e.g. J\"ager et al. 2004) and is reasonable for
the achieved spectral resolution (and a slit width of 1\arcsec).

From this velocity field we can further extract a rotation curve by
modelling a slit which is placed over the major axis of the galaxy
with a typical slit width of 1\arcsec, as detailed in Kronberger et
al. (2006).

\section{Results}

In the following we investigate observable projections of the full
3D velocity field of the simulation, i.e. the rotation curves and
the 2D velocity fields. Special emphasis is put on the question how
the given distortions of the velocity field appear at intermediate
redshift.

We quantify the degree of asymmetry of the RC shape, following an
approach used in Dale et al. (2001) and subsequently for simulated
galaxies in Kronberger et al. (2006). The area between the
kinematically folded approaching and receding halves is divided by
the average area under the RC:

\begin{eqnarray}\label{assymeq}
\textrm{Asymmetry}&=&\sum \frac{||V(r)|-|V(-r)||}{\sqrt{\sigma^2(r)
+\sigma^2(-r)}} \nonumber\\ &\times& \left[\frac{1}{2} \sum
\frac{|V(r)|+|V(-r)|}{\sqrt{\sigma^2(r)+\sigma^2(-r)}}\right]^{-1}.
\end{eqnarray}

Here $V(r)$ is the velocity and $\sigma(r)$ is the uncertainty of
the rotational velocity at position $r$. As in Kronberger et al.
(2006) we adopt for $\sigma(r)$ a value of 20 km/s, which is
typically the minimum uncertainty in observations of distant
galaxies due to the limited spectral resolution.

In order to quantify the distortions and to interpret the partly
complex structures in the 2D velocity fields we use as in Kronberger
et al. (2007) the kinemetry package of Krajnovi\'{c} et al. (2006).
This method is based on the assumption that the mean velocity along
best fitting ellipses is reproduced by a simple cosine law, i.e.

\begin{equation}
V(a,\Psi)=V_0+V_c(a)cos{\Psi},
\end{equation}

\noindent where $a$ is the length of semi-major axis of the ellipse,
$\Psi$ for discs is the azimuthal angle measured from the major axis
in the plane of the galaxy. The position angle $\Gamma$ and the
axial ratio ($q=b/a$) of the ellipses are calculated as a function
of radius from the galactic centre. Deviations from the cosine law
are measured using an harmonic expansion along the ellipses, i.e.

\begin{equation}
V(a,\Psi)=\sum_{n=1}^{N}k_n(a)cos[n(\Psi-\phi_n(a))],
\end{equation}

\noindent where $\phi$ is the phase coefficient. Higher order
Fourier terms and radial changes of $\Gamma$ or $q$ quantize
deviations in the velocity field from a simple rotation. Sometimes a
simple sine correction for the inclination of the disc is applied in
observations. As this correction simply alters the absolute value of
the velocities, we do not apply it to our model velocity fields. All
VFs were extracted at the same inclination i$=$35$^{\circ}$.

\begin{figure}
\begin{center}
{\includegraphics[width=\columnwidth]{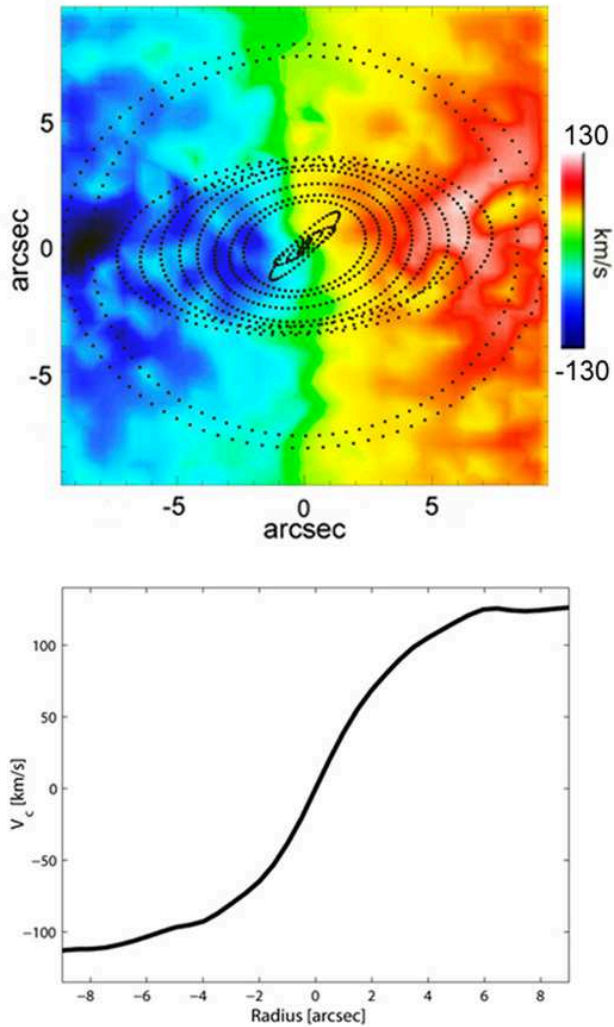}} \caption{2D
velocity field and long-slit rotation curve of an undisturbed model
galaxy as seen at redshift $z=0.1$. The figure of the 2D velocity
field is taken from Kronberger et al. (2007). Overlayed as contours
are the best fitting ellipses from the kinemetry analysis.}
\label{undisturbed}
\end{center}
\end{figure}

In Fig. \ref{undisturbed} we present the 2D velocity field and a
long-slit RC from an undisturbed model galaxy as seen at redshift
$z=0.1$. For the appearance of the VF at different redshifts we
refer to Kronberger et al. (2007).

\subsection{Rotation curves of ram-pressure stripped galaxies}

\begin{figure}
\begin{center}
{\includegraphics[width=\columnwidth]{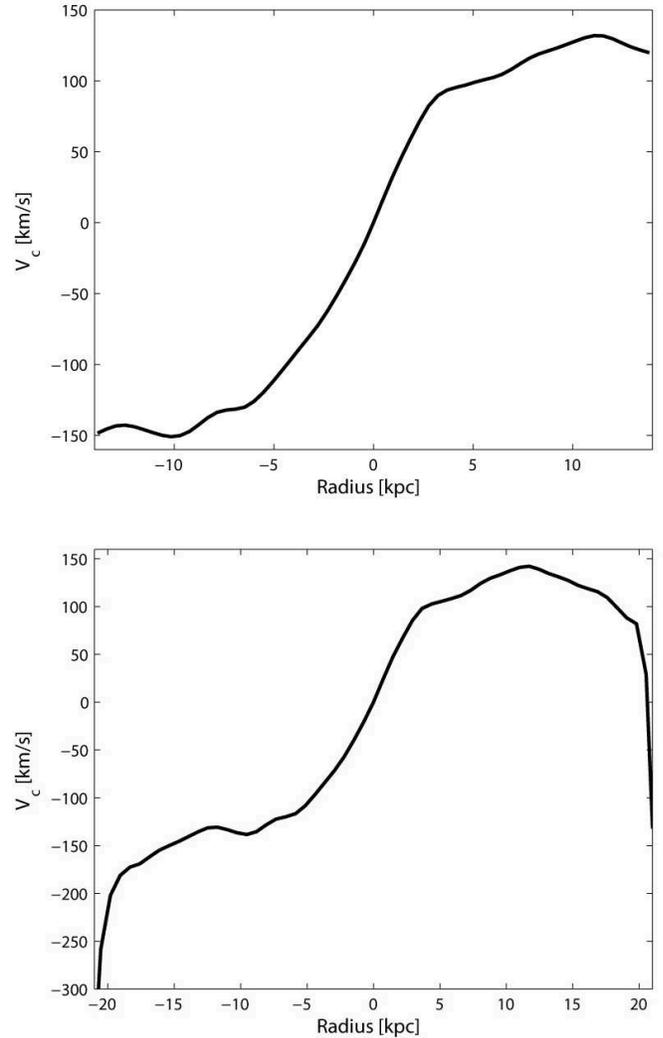}}
\caption{Rotation curve of a model galaxy after 50 Myr of ram
pressure acting face-on. The RC is assumed to be observable out to 4
(top panel) and 6 (lower panel) times the radial disc scale length,
respectively. In the top panel, the RC is similar to the initial
undisturbed configuration. In the lower panel signatures of
ram-pressure stripping are clearly visible in the outer parts. Note
that the 'direction' of the distortion depends on the movement of
the galaxy with respect to the observer. If the galaxy was moving
away from the observer, the rotational velocities in the outer parts
of the RC would go to higher \textit{positive} values.}
\label{rc5wide}
\end{center}
\end{figure}

First we study the RC of a model galaxy moving \textit{face-on}
through the ambient medium. After 100 Myr of interaction a clear
stripping radius is visible, which stays roughly constant over time
and agrees with analytical estimates (Kronberger et al. 2008). The
stripping radius in ram-pressure affected galaxies is defined, as
the distance from the galactic centre, outside which the
inter-stellar medium (ISM) cannot be prevented from being stripped
by the galactic gravitational potential. We extract rotation curves
from the galaxy before the clear emergence of a stripping radius,
i.e. in the early phases of stripping (after $\sim$ 50 Myr), and
thereafter (after $\sim$ 100 Myr). The RC of the isolated model
galaxy has the typical shape, rising in the inner and flattening in
the outer parts. After 50 Myr of acting ram pressure first effects
of the external pressure are visible in the distribution of the gas.
The strength of the disturbances depends on the local gravitational
potential. They are strongest at large galactic radii, where the
restoring force acting against ram pressure is weakest. Whether
these distortions are visible in the rotation curve depends on the
radius to which the RC can be measured. We assume that the
rotational velocity can be determined accurately out to 4 times the
radial disc scale length of the galaxy. This is a typical value
obtainable at intermediate redshift (e.g. Ziegler et al. 2003). In
the top panel of Fig. \ref{rc5wide} we present the corresponding RC,
which does not show strong peculiarities (asymmetry measure
$\sim$6\%).

Local measurements extend to much larger radii. We consider the case
where the rotational velocity can be measured out to 6 times the
radial disc scale length. Already after 50 Myr, the gas at such
large radii gets a dominant velocity component antiparallel to the
galaxies' direction of movement, i.e. away from the galactic disc.
How this change of the velocity vector in the stripping region
translates to a change in the RC of the galaxy is shown in the lower
panel of Fig. \ref{rc5wide}. In the outer parts of the RC, a
turn-over of the rotational velocity on both sides to higher
negative values can be observed. This is a consequence of gas moving
away from the galactic disc due to the ram-pressure. The asymmetry
measure of this disturbed RC is $\sim$ 22\%, which is low compared
to distortions caused by tidal interactions (see Kronberger et al.
2007). Note that the 'direction' of the distortion depends on the
movement of the galaxy with respect to the observer. If the galaxy
was moving away from the observer, the rotational velocities in the
outer parts of the RC would go to higher \textit{positive} values.

After 100 Myr the stripping radius at r$\sim$12 kpc is clearly
visible. Hence also the RC distortion is now visible in this region
of the disc, and also observable at intermediate redshift. The RC
for this time step is presented in Fig. \ref{RC10_50} (top panel).
There is a dependence of the RC shape on the inclination of the
galactic disc but the important characteristics of the RC remain
observable for most inclinations. These characteristic features in
the outer parts of the RC are less prominent if the galaxy is
observed almost edge-on. There the gas velocity is dominated by a
component perpendicular to the galactic disc due to the acceleration
by the ICM wind. If the galaxy is seen edge-on, only the altered
disc velocity component is observed. The dependence on the viewing
angle is not as strong as for tidal interactions (see Kronberger et
al. 2006) as ram-pressure stripping is an axisymmetric distortion.
The characteristic features are observable for all viewing angles.

\begin{figure}
\begin{center}
{\includegraphics[width=\columnwidth]{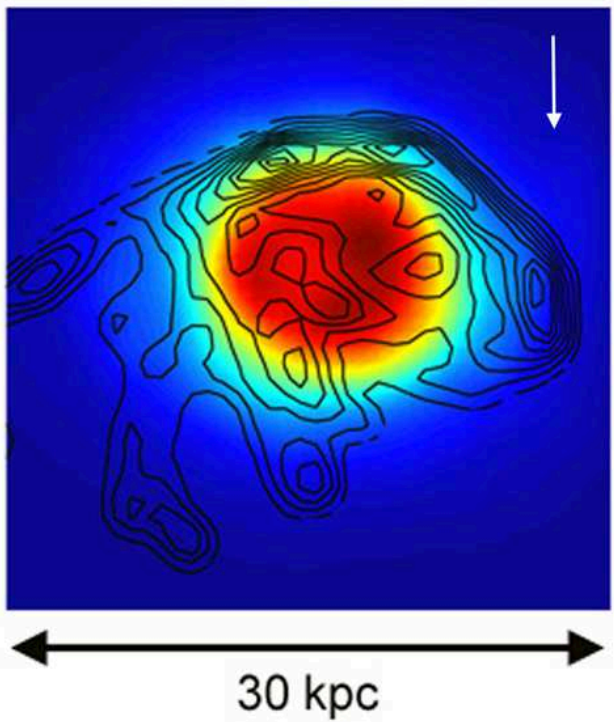}} \caption{The
gas distribution is overlayed as contours on the smoothed stellar
distribution after 500 Myr of ram pressure acting edge-on. The white
arrow in the top right corner of the figure indicates the ICM wind
direction. The galaxy rotates counter-clockwise.}
\label{contours_eo}
\end{center}
\end{figure}

Next we investigate the RC of a model galaxy flying \textit{edge-on}
through the ambient medium. In this case the perturbation acts in
the plane of the disc. The gaseous disc gets strongly compressed,
which alters the radial pressure gradients and hence the rotational
velocity of the gas. We find stronger distortions than in the
face-on case, especially a mismatch between kinematic centre and the
centre of the stellar disc is observable. The offset of the gas
kinematic centre is $\approx$ 2 kpc after 500 Myr of ram pressure
acting. In Fig. \ref{contours_eo} we show the contours of the gas
distribution overlayed on the smoothed stellar distribution after
500 Myr of ram-pressure acting (we chose this evolved time step,
compared to the face-on case, as the distortions are best visible
there. Before, the distortions are qualitatively the same, but less
prominent). In contrast to the face-on stripping, there is no
distinct centre or symmetry visible in the gas distribution anymore.
In the rotation curve, there are basically two features visible. As
the kinematic centre does not coincide with the optical centre of
the galaxy anymore, the RC is slightly more asymmetric than in the
case of face-on ram pressure (asymmetry measure $\sim$30\%).
Secondly, the RC is declining in the outer part on both sides. Note
that this feature is in this case only visible in the RC of the gas,
as only this component is affected by the ram pressure of the
ambient medium. The collisionless components of the galaxy (i.e. the
stars and the dark matter) are not or due to the changing gas
density distribution only mildly affected. If a declining RC is also
observed in the stellar RC, a tidal interaction affecting also the
stars and the dark matter is most likely the cause. There is a
dependence on the viewing angle and the inclination of the disc, not
strongly affecting, however, the main feature of the disturbed RC,
namely the declining outer parts.

\begin{figure}
\begin{center}
{\includegraphics[width=\columnwidth]{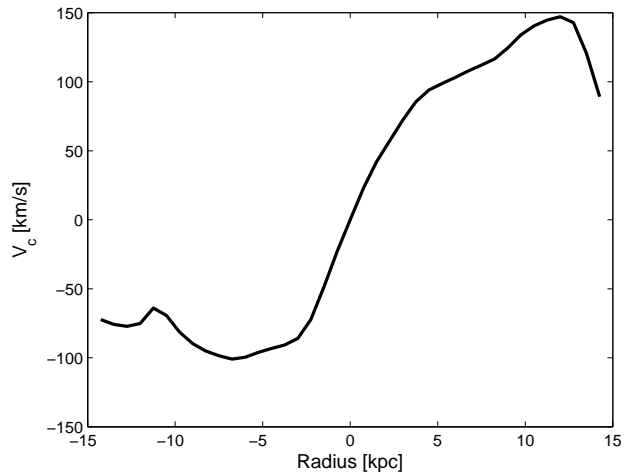}}
\caption{Rotation curve of a model galaxy after 500 Myr of ram
pressure acting edge-on.} \label{RC50_eo}
\end{center}
\end{figure}

For better visual judgement of the asymmetry of the RC, both its
sides are sometimes plotted on one side. In Fig. \ref{RC10_50} we
present the RCs in such a way for the model galaxy after 100 Myr of
face-on ram pressure and after 500 Myr of edge-on ram pressure,
respectively.

In both cases, edge-on and face-on, the long-slit RC cannot be used
to estimate parameters for a Tully--Fisher study. The kinematics of
the gas in the galaxy is affected by ram-pressure stripping. The
unaffected stellar kinematics could be used instead to estimate the
maximum rotational velocity $V_{max}$. However, the luminosity of
the galaxy is also affected by ram-pressure stripping. In a recent
work, we found an enhancement of the star-formation rate due to ram
pressure (Kronberger et al. 2008). A subsequent brightening of the
ram-pressure affected galaxy can cause an increased scatter in the
Tully--Fisher relation.

\begin{figure}
\begin{center}
{\includegraphics[width=\columnwidth]{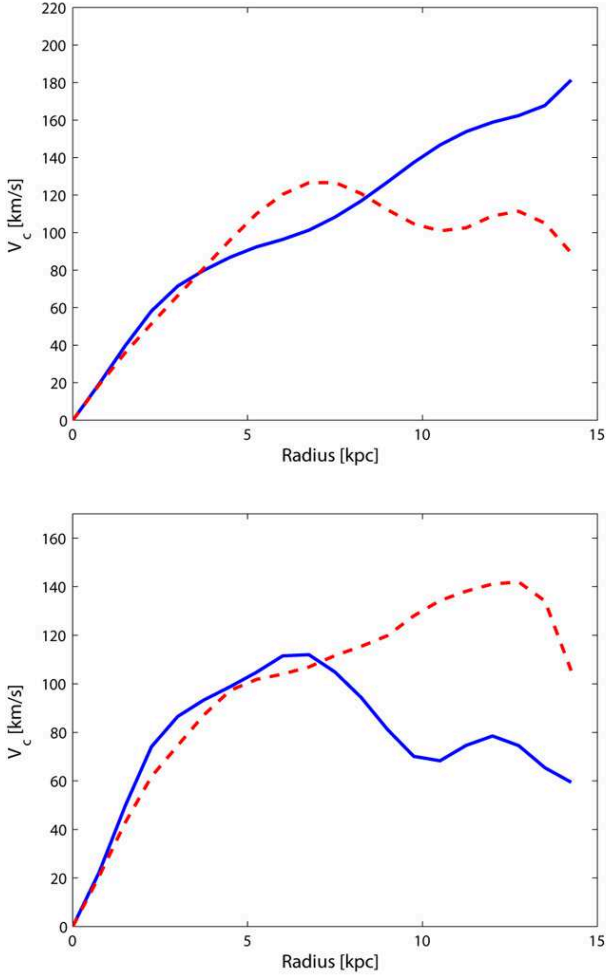}}
\caption{Rotation curve of a model galaxy after 100 Myr of ram
pressure acting face-on (top panel) and after 500 Myr of ram
pressure acting edge-on (bottom panel). Both sides of the RC are
plotted on one side for better judgement of the asymmetry. The
blue-shifted part of the RC is plotted as blue solid line, while the
redshifted part is plotted as dashed red line.} \label{RC10_50}
\end{center}
\end{figure}

\begin{figure}
\begin{center}
{\includegraphics[width=\columnwidth]{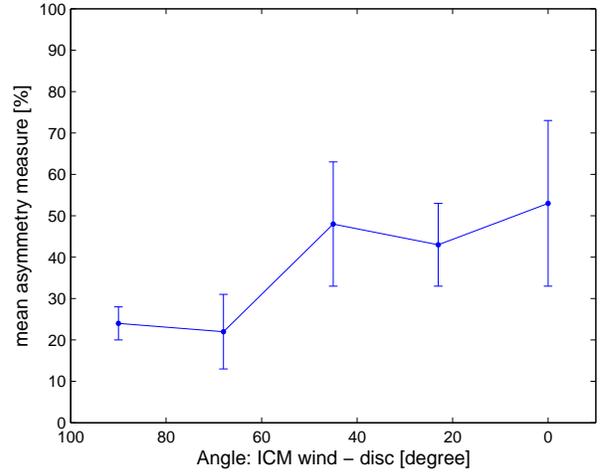}}
\caption{Asymmetry measure as a function of the angle between the
disc and the ICM wind direction (0$^{\circ}$ is edge-on and
90$^{\circ}$ is face-on ram pressure). The asymmetry measure was
averaged over several viewing angles. The error bars indicate the
standard deviation of the mean.} \label{assym}
\end{center}
\end{figure}

In Fig. \ref{assym} we study how the asymmetry of the VF depends on
the angle between the ICM wind direction and the plane of the disc.
The asymmetry measure depends on the viewing angle and the
inclination, so we averaged over three different lines-of-sight and
two observed disc inclinations. A clear step in the asymmetry
measure can be seen at 45$^{\circ}$, where the asymmetry increases
significantly. The scatter, however, increases as well, which makes
it difficult to distinguish further between different ICM wind
directions with an observed VF asymmetry. A high VF asymmetry is
nevertheless a good hint, for a small angle between the ICM wind
direction and the plane of the disc.

\subsection{2D velocity fields of ram-pressure stripped galaxies}

With the help of 2D velocity fields the nature of the interaction
becomes more accessible than with simple rotation curves from
long-slit spectroscopy. Quantitative analyses on the kinematic axis,
the kinematic centre and more detailed analyses such as harmonic
expansion, which is, for example, used by the kinemetry package of
Krajnovi\'{c} et al. (2006), offer additional possibilities to
identify distortions in a VF. We have already applied the kinemetry
method to tidally disturbed simulated systems in Kronberger et al.
(2007).

\begin{figure}
\begin{center}
{\includegraphics[width=\columnwidth]{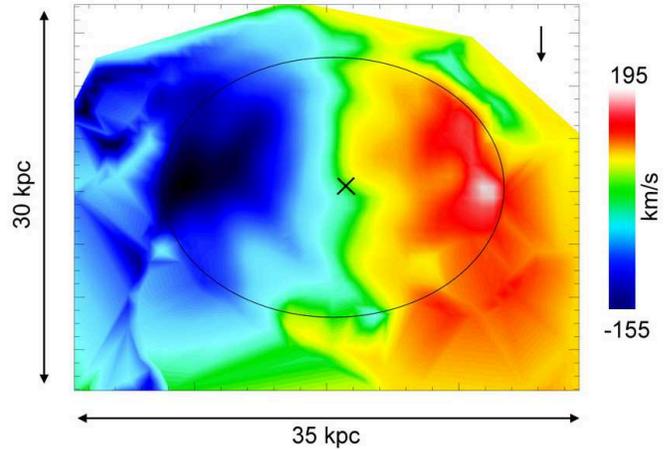}} \caption{2D
velocity field of a face-on ram-pressure affected galaxy after 250
Myr of ram pressure acting. The almost undisturbed inner part of the
velocity field is highlighted by an ellipse, whose semi-major axis
roughly corresponds to the stripping radius ($\sim$12 kpc as
estimated in Kronberger et al. 2008). The centre of the stellar disc
is indicated by the black cross. The black arrow in the top right
corner indicates the projected ICM wind direction. Note that the ICM
wind acts perpendicular to the plane of the disc.} \label{VF26}
\end{center}
\end{figure}

\begin{figure}
\begin{center}
{\includegraphics[width=\columnwidth]{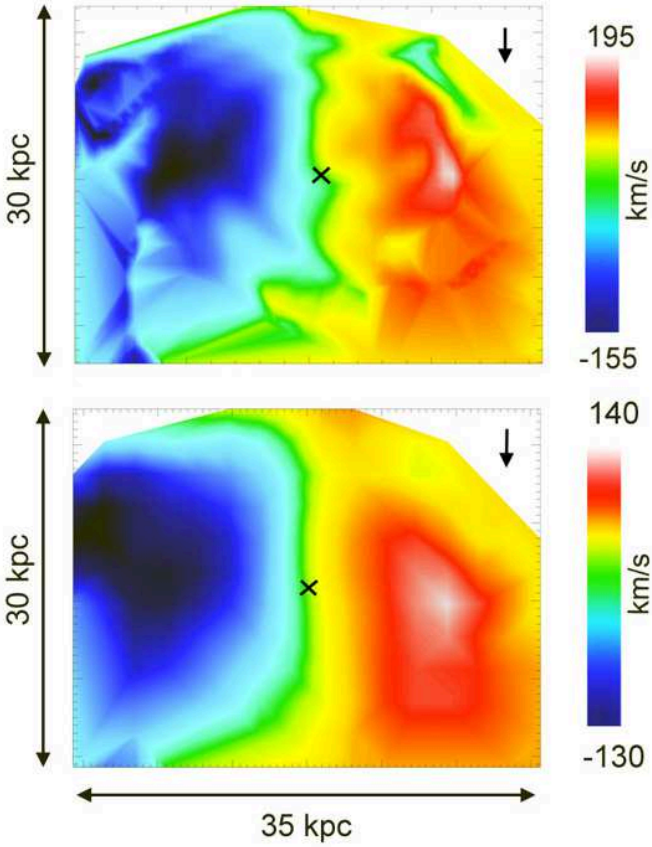}} \caption{The
2D velocity field of a face-on ram-pressure affected galaxy after
250 Myr of ram pressure acting as seen at redhift z=0.1 (top panel)
and z=0.4 (lower panel), respectively. The centre of the stellar
disc is indicated by the black cross. The black arrow in the top
right corner of each panel indicates the projected ICM wind
direction. Note that the ICM wind acts perpendicular to the plane of
the disc.} \label{VF26_z}
\end{center}
\end{figure}

In Fig. \ref{VF26} we show such a 2D velocity field of a face-on
ram-pressure affected galaxy after 250 Myr of ram pressure acting.
(The distortions at this time step are qualitatively equal to those
present after 100 Myr, but more prominent. Therefore we chose this
particular time step, without loss of generality). The distortions
are not as severe as those caused by tidal interactions, which were
presented in Kronberger et al. (2007). An almost undisturbed inner
part of the velocity field is clearly visible, which we indicated
with an ellipse. The semi-major axis of this ellipse corresponds
roughly to the stripping radius (12 kpc) of this model system as
estimated in Kronberger et al. (2008). Beyond this radius the VF
appears disturbed. This picture of an undisturbed inner part and a
severely affected outer part beyond the stripping radius also agrees
with the rotation curve of the system presented in the previous
section (c.f. Fig. \ref{rc5wide}). As for the rotation curve, the VF
appearance depends on the inclination. The characteristic features
become less prominent when the galaxy is observed close to edge-on.
In Fig. \ref{VF26_z} we study the appearance of this velocity field
when put at intermediate redshift (the same investigation as
performed in Kronberger et al. (2007) for tidally disturbed velocity
fields).

\begin{figure}
\begin{center}
{\includegraphics[width=\columnwidth]{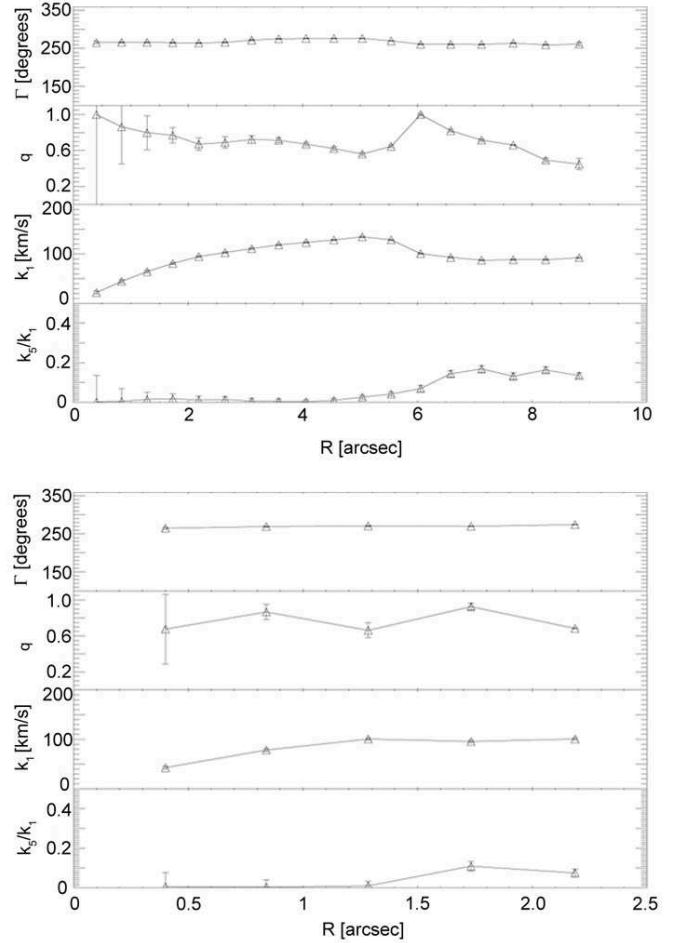}}
\caption{Radial profiles of the kinemetric properties, calculated
using the kinemetry programme for the model galaxy after 250 Myr of
face-on ram pressure acting, for redshift $z=0.1$ (top) and $z=0.4$
(bottom). The position angle $\Gamma$ and the flattening $q$ of the
best fitting ellipses as well as the first and the fifth order
Fourier terms $k_1$ and $k_5$ are plotted as a function of radius.}
\label{plots26}
\end{center}
\end{figure}

The VF appears smooth and widely undisturbed at $z=0.4$, as the
irregularities are smeared out, mainly due to seeing (therefore the
use of integral field spectrographs in combination with adaptive
optics is desirable). The radial profiles of the kinemetric
properties, which we calculated using the kinemetry programme of
Krajnovi\'{c} et al. (2006), are presented in Fig. \ref{plots26} for
the model galaxy after 250 Myr of face-on ram pressure acting. The
position angle $\Gamma$ and the flattening $q$ of the best fitting
ellipses as well as the first and the fifth order Fourier terms
$k_1$ and $k_5$ are plotted as a function of radius for redshift
$z=0.1$ (top) and $z=0.4$ (bottom). In the radial behaviour of these
quantities the undisturbed inner part of the velocity field and the
distorted outer part are visible. At a radius of $\sim$ 5.5
$\arcsec$ (at z=0.1, corresponding to $\sim$ 10 kpc) a clear
increase in the ratio of $k_5$ and $k_1$ can be seen. This ratio
represents complex, kinematically separate components in the
velocity field. At the same radius the flattening of the ellipses
and the first order Fourier term $k_1$ also show significant
variations, which are, however, lower than in the case of tidally
induced distortions (c.f. corresponding figures in Kronberger et al.
2007). At a redshift of $z=0.4$, the variations of the radial
kinemetric properties get less prominent. A clear identification of
ram pressure affected galaxies is especially difficult as at such
redshifts often only the inner part of the velocity field is
observable. In the outer regions of the disc some irregularities in
the velocity field remain visible also at intermediate redshift.
They are observable as increase in the ratio of $k_5$ and $k_1$.

\begin{figure}
\begin{center}
{\includegraphics[width=\columnwidth]{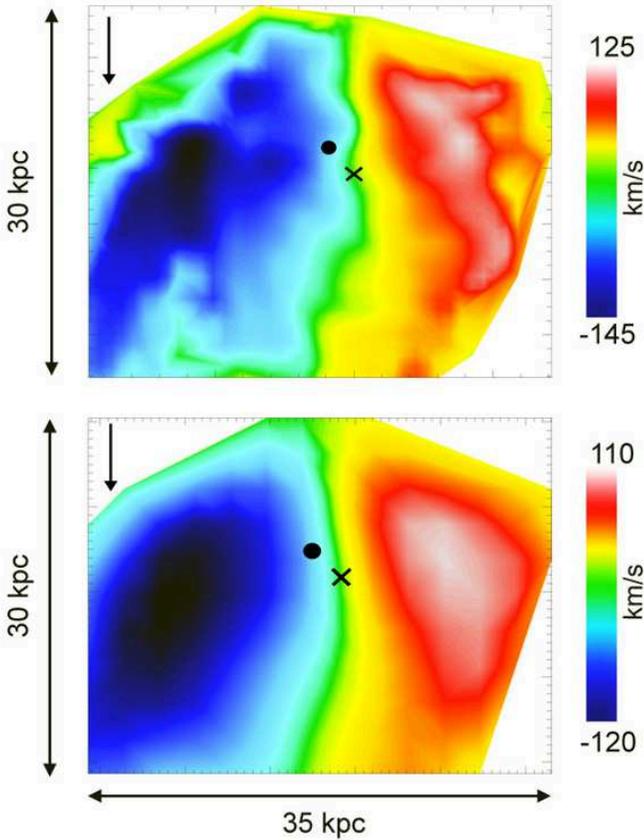}} \caption{2D
velocity field of an edge-on ram-pressure affected galaxy after 400
Myr of ram pressure acting as seen at redhift z=0.1 (top panel) and
z=0.4 (lower panel), respectively. The black arrow in the top left
of each panel indicates the ICM wind direction and the cross and the
circle indicate the kinematic and stellar disc centre,
respectively.} \label{VF42_eo}
\end{center}
\end{figure}

\begin{figure}
\begin{center}
{\includegraphics[width=\columnwidth]{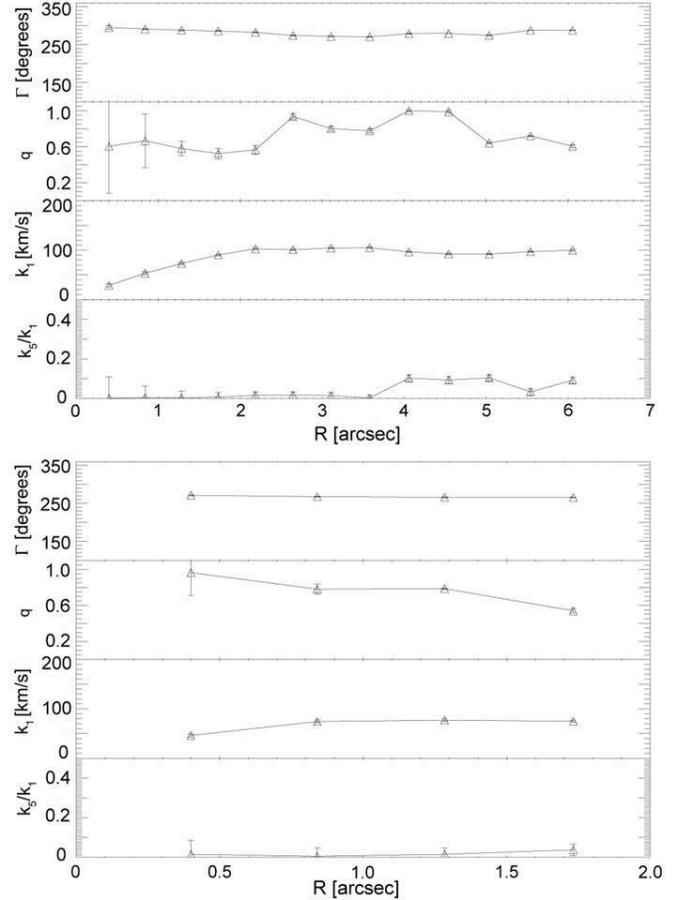}}
\caption{Radial profiles of the kinemetric properties, calculated
using the kinemetry programme for the model galaxy after 400 Myr of
edge-on ram pressure acting, for redshift $z=0.1$ (top) and $z=0.4$
(bottom). The position angle $\Gamma$ and the flattening $q$ of the
best fitting ellipses as well as the first and the fifth order
Fourier terms $k_1$ and $k_5$ are plotted as a function of radius.}
\label{plots_eo}
\end{center}
\end{figure}

In Fig. \ref{VF42_eo} we present the 2D velocity field of an edge-on
ram-pressure affected galaxy after 400 Myr of ram pressure acting as
seen at redhift z=0.1 (top panel) and z=0.4 (lower panel),
respectively. The mismatch between kinematic centre and the centre
of the stellar disc is the strongest effect of edge-on ram pressure.
Already by visual assessment of the VF presented in Fig.
\ref{VF42_eo} several distortions are visible also in the inner part
of the VF. These irregularities are, however, completely smeared out
at redshift $z=0.4$. Also the radial profiles of the kinemetric
properties, presented in Fig. \ref{plots_eo} reflect this behaviour.
At $z=0.1$ variations of the flattening parameter $q$ and of the
ratio of $k_5$ and $k_1$ are already visible at smaller radii. The
amplitude of these variations is however small. To study the spatial
distribution of the cold gas seems therefore to be still the best
way to identify ram pressure acting in a galaxy. The influence on
the velocity field is relatively low, as the rotational velocity
depends mostly on the gradients of the gravitational potential and
the gas accounts only for a small mass fraction.

\section{Discussion}

Ram-pressure stripping has been modelled numerically using different
techniques. All these approaches use some simplifications due to the
complexity of the multi-phase, multi-scale gas physics present in
the inter-stellar medium of galaxies. One method is to treat the ISM
hydrodynamically either with an Eulerian code (e.g. Roediger \&
Br\"uggen 2007) or using SPH (e.g. J{\'a}chym et al. 2007). SPH has
the advantage that it can be easily combined with collisionless
N-body simulations for the dark matter and stellar component. In the
setup used for this work the collisionless component and the gas are
treated fully self-consistently.  An alternative method is to model
the ISM with particles which can have inelastic collisions,
so-called `sticky particles´ (e.g. Vollmer et al. 2006). The ICM ram
pressure is then modelled as additional acceleration on particles,
which are not shielded by others. All these methods have their
advantages and restrictions. If the ISM is considered as continuous
fluid the hydrodynamic prescription yields a better model than
sticky particles. The latter method is more suitable for
non-continuous systems (e.g. Vollmer et al. 2006). The SPH method
has intrinsic shortcomings in the resolution of instabilities and
turbulence (e.g. Agertz et al. 2007). In systems affected by
ram-pressure stripping shear flows are present, which can lead to
the development of Kelvin-Helmholtz instabilities (e.g. Roediger et
al. 2006). These are not resolved in our setup. The Kelvin-Helmholtz
instabilities affect mostly the stripped material in the wake, where
the shear flow is strong and the gravitational force is weak
(Roediger et al. 2006). The effect for the bulk motion of the gas in
the plane of the disc is probably small. A similar study as the one
presented in the present work should be carried out with a Eulerian
code in a future work to investigate the importance of hydrodynamic
instabilities in this issue. The huge advantage of our setup is that
we treat gas, dark matter and the stellar component fully
self-consistently.

Some of the results presented in this work were also found in
previous studies. Roediger \& Br\"uggen (2007), for example, found
that the gas distribution of ram-pressure affected galaxies is more
asymmetric for edge-on ram pressure. Consistently we find this trend
also in the velocity field. The mismatch between the kinematic and
the stellar disc centre is caused by two effects. Ram pressure
affects the whole disc. The gas gets dislocated where the restoring
gravitational force is too low. This ram-pressure pushing (Roediger
\& Br\"uggen 2007) leads also to a central density enhancement and
consequently to an increased central star formation (Kronberger et
al. 2008; Vollmer et al. 2001). Within our model this enhanced star
formation is self-consistently treated and induces an increased
galactic wind. These feedback mechanisms, which do affect the gas
kinematics due to modified density and pressure gradients, were
neglected in previous studies and are one of the strengths of the
present work. With an offset of $\approx$2 kpc, the effect is,
however, relatively small.

In a series of papers Vollmer et al. use the velocity field
information to fit models of ram-pressure affected galaxies to HI
observations of nearby cluster galaxies (e.g. Vollmer et al., 2008).
They identify characteristic features in the VFs of the observed
systems and study in which configuration the model can reproduce
these features. In Vollmer et al. (2008) the velocity field of an
almost edge-on stripped galaxy is shown. The VF is quite regular as
the galaxy is in an early phase of stripping. Thus a direct
comparison to our models is difficult. Even for edge-on stripping
the VF distortions are moderate in our models. This is also true for
the simulations of Vollmer et al. (2008) (compare e.g. their Fig.
20). Also in Vollmer et al. (2006) the VF appears rather regular.
The VF of NGC 4569 presented in Vollmer et al. (2004) shows an
asymmetric velocity field with an angle between orbital and disc
plane of 35$^{\circ}$. This is consistent with our finding that the
velocity field becomes asymmetric for angles $<$45$^{\circ}$.

More detailed comparisons with observed optical velocity fields and
rotation curves from the sample of Ziegler et al. (2007) will be
subject of an upcoming work.

\section{Summary and conclusions}

We have investigated 2D velocity fields and rotation curves of the
gas in ram-pressure affected spiral galaxies using N-body/SPH
simulations.

\begin{itemize}
\item We found that the appearance of distortions of the gaseous disc
due to ram-pressure stripping depends on the direction of the acting
ram-pressure. In our investigation we studied the two extreme cases
of face-on and edge-on ram pressure. In both cases, the velocity
fields of the ram-pressure affected galaxies show characteristic
features.

\item In general, the distortions of the gas velocity field are significantly lower
than in the case of distortions induced by tidal interactions. The
reason is that the rotational velocity depends mostly on the
gravitational potential and the gas accounts only for a small mass
fraction.

\item In the case of face-on acting ram pressure distortions appear mainly in the
outer parts of the galaxy. The separation between undisturbed inner
and symmetrically disturbed outer part is visible in both the
rotation curve and the 2D velocity field.

\item In the case of ram pressure acting edge-on we find a significant mismatch between
kinematic centre and the centre of the stellar disc. The rotation
curve is asymmetric and the outer parts of the RC are declining. The
2D velocity field shows signatures of the interaction also in the
inner part of the disc.

\item At an angle of 45$^{\circ}$ between the ICM wind direction and the disc,
the velocity field asymmetry increases significantly with respect to
smaller angles. The scatter is, however, increasing towards the
edge-on wind too. Therefore it seems difficult to distinguish
further between different ICM wind directions purely with the help
of the VF asymmetry.

\item The collisionless stellar disc is not affected by ram pressure and hence
shows no disturbed kinematics. The characteristic features in the
gas kinematics presented here, together with an absence of
distortions in the stellar kinematics are therefore a strong
indication for the presence of ram-pressure stripping acting on a
cluster galaxy.

\end{itemize}

\begin{acknowledgements}
We thank the anonymous referee for fruitful comments which helped to
improve the quality of the paper. The authors thank Volker Springel
for providing them with GADGET2 and his initial-conditions generator
and Davor Krajnovi\'{c} for his Kinemetry software. Thomas
Kronberger is a recipient of a DOC fellowship of the Austrian
Academy of Sciences. The authors further acknowledge the
UniInfrastrukturprogramm des BMWF Forschungsprojekt Konsortium
Hochleistungsrechnen, the ESO Mobilit\"atsstipendien des BMWF
(Austria), the Austrian Science Foundation (FWF) through grants
P18523-N16 and P19300-N16, the German Science Foundation (DFG)
through Grant number Zi 663/6-1, the Volkswagen Foundation (I/76
520), and the Tiroler Wissenschaftsfonds (Gef\"ordert aus Mitteln
des vom Land Tirol eingerichteten Wissenschaftsfonds). We further
thank Chiara Ferrari for fruitful discussion.
\end{acknowledgements}

\end{document}